%% file: main.tex
\documentclass[journal]{IEEEtran}
\usepackage{graphicx} % Required for inserting images
\usepackage{tikz} % Pictures
\usepackage{listings}
\usepackage{enumitem}
\usepackage{graphicx}
\usepackage{multicol, latexsym}
\usepackage{blindtext}
\usepackage{caption}
\usepackage{longtable}
\usepackage{csquotes}
\usepackage{amsfonts}
\usepackage{amsmath}
\usepackage{amsthm}
\usepackage{amssymb}
\usepackage{algorithm}
\usepackage{tabularx}
\usepackage{capt-of,lipsum} % dedicated caption for equation
\usepackage{balance}
\usepackage{subfigure} 
\usepackage{listings}
\usepackage{comment}
\usepackage{diagbox}
\usepackage{pdflscape}
\usepackage{booktabs}
\usepackage{makecell}
\usepackage{lipsum}
\usepackage{adjustbox}

\newcommand{\ie}{\textit{i}.\textit{e}.,\ }
\newcommand{\eg}{\textit{e}.\textit{g}.,\ }
\newcommand{\cf}{\textit{c}\textit{f}.\ }
\newcommand\1{\textit{(i)}}
\newcommand\2{\textit{(ii)}}
\newcommand\3{\textit{(iii)}}

\newcommand\approach{\textit{QBER}}

\begin{document}

\title{\approach: Quantifying Cyber Risks for \\ Strategic Decisions}
\author{
    % Muriel Figueredo Franco
    \IEEEauthorblockN{Muriel Figueredo Franco, Aiatur Rahaman Mullick, Santosh Jha\\}
    \IEEEauthorblockA{Zeron – The Single Point of Truth for Cybersecurity  \\
    Mumbai, India \\
        Email: [muriel, aiatur, santosh]@zeron.one\\
        % Email: mffranco@inf.ufrgs.br
    } \and

}

\maketitle

\begin{abstract}
Quantifying cyber risks is essential for organizations to grasp their vulnerability to threats and make informed decisions. However, current approaches still need to work on blending economic viewpoints to provide insightful analysis. To bridge this gap, we introduce \approach{} approach to offer decision-makers measurable risk metrics. The \approach{} evaluates losses from cyberattacks, performs detailed risk analyses based on existing cybersecurity measures, and provides thorough cost assessments. Our contributions involve outlining cyberattack probabilities and risks, identifying Technical, Economic, and Legal (TEL) impacts, creating a model to gauge impacts, suggesting risk mitigation strategies, and examining trends and challenges in implementing widespread Cyber Risk Quantification (CRQ). The \approach{} approach serves as a guided approach for organizations to assess risks and strategically invest in cybersecurity.
\end{abstract}

\IEEEpeerreviewmaketitle


\input{sections/introduction}
\input{sections/relatedwork}
\input{sections/approach}

\input{sections/trends}
\input{sections/conclusions}
%\section*{Acknowledgments} % In case there are some people and team to thank.
\balance
\bibliographystyle{IEEETranCustomized}
\bibliography{references.bib}
\vspace{-0.3cm}
\noindent \small{\\All links provided above were last accessed on May 2024.}
\newpage
\section*{Biography}
\vspace{-2.0cm}
\begin{IEEEbiography}[{\includegraphics
[width=1in,height=1.25in,clip,keepaspectratio]{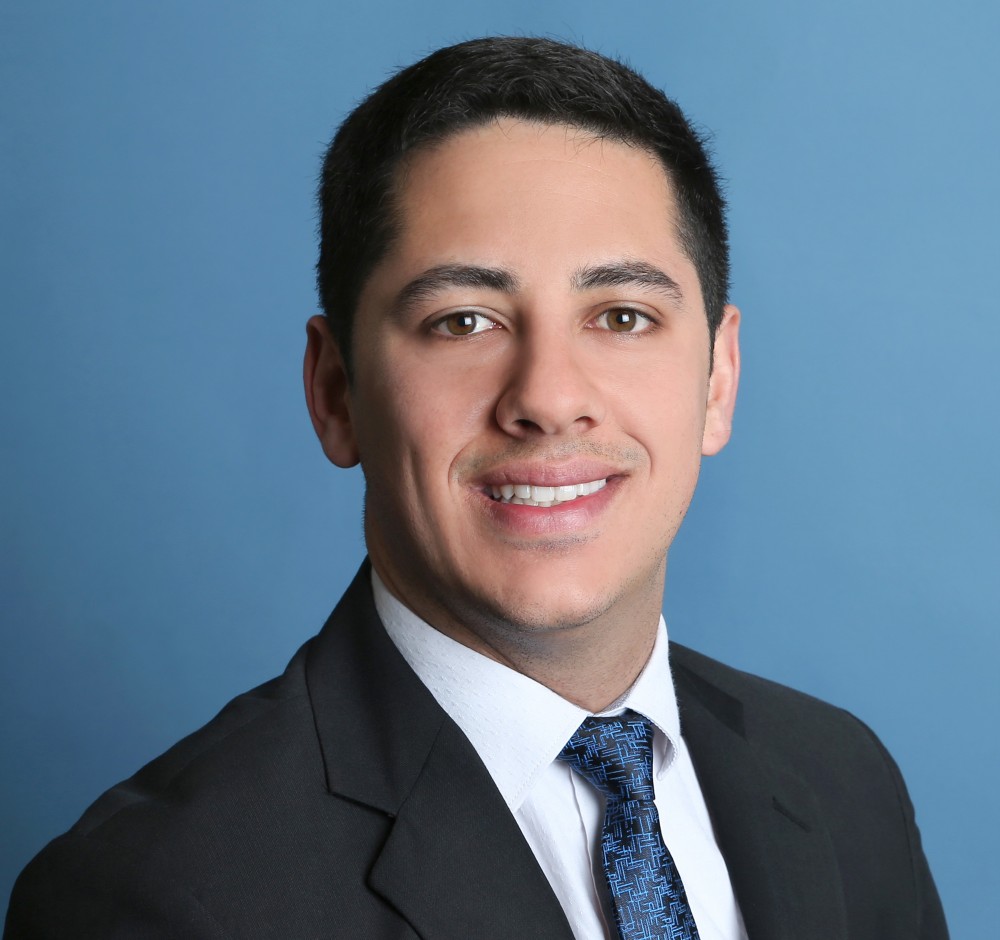}}]
{Dr. Muriel F. Franco} is a Senior Research Consultant at Zeron and Postdoctoral Researcher at the Computer Networks Group of Federal University of the Rio Grande do Sul (UFRGS). In his research career, he participated in different multidisciplinary projects within teams of networking, cybersecurity, and economic experts, with over 70 research papers and patents co-authored. Muriel holds a PhD (Summa cum laude) from 2023 in Computer Science from the University of Zurich UZH, Switzerland, MBA from 2022 in Project Management from the University of São Paulo (USP), Brazil, MSc from 2017 in Computer Science from Federal University of Rio Grande do Sul (UFRGS), Brazil, and received a BSc from 2014 in Computer Science from the Federal University of Pelotas (UFPEL), Brazil. His research interests include cybersecurity, network management, risk management, and communication systems. 
\end{IEEEbiography}
\vspace{-1cm}

\begin{IEEEbiography}[{\includegraphics
[width=1in,height=1.25in,clip,keepaspectratio]{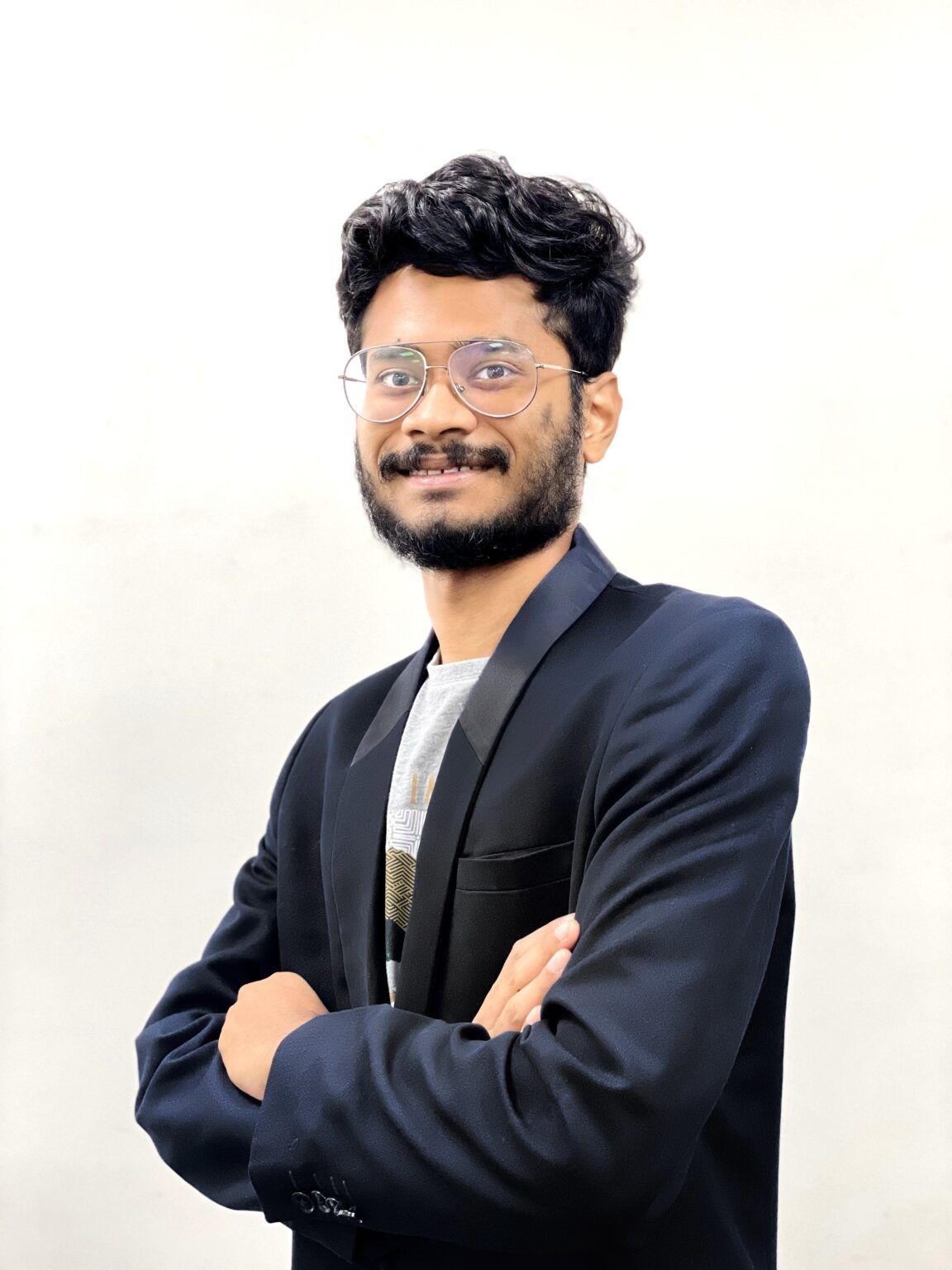}}]
{Aiatur R. Mullick} is an experienced cybersecurity professional at Zeron who has the expertise of effectively guide enterprises in defining priorities for their cyber risks. Mullick's expertise rests in engaging with organisations to establish comprehensive risk registers and implement tailored risk mitigation plans. With considerable expertise in implementing controls for various regulations and standards, Aiatur delivers a feasible, results-oriented approach to enhancing enterprise cybersecurity posture.
\end{IEEEbiography}
% Add other authors here
\vspace{-1cm}

\begin{IEEEbiography}[{\includegraphics
[width=1in,height=1.25in,clip,keepaspectratio]{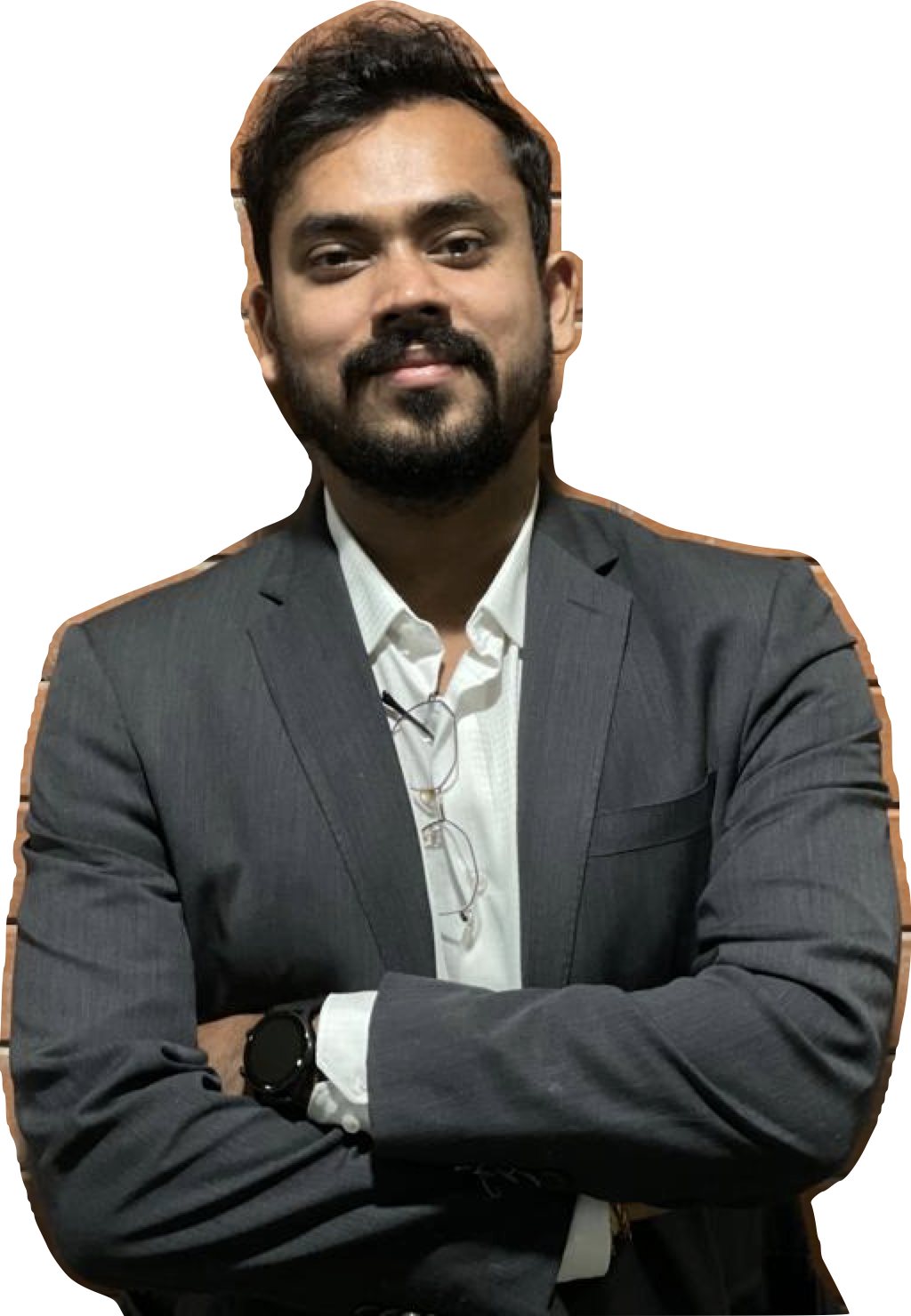}}]
{Santosh Jha} is a vivid cybersecurity researcher and Chief Technology Officer at Zeron. He is a Certified Ethical Hacker and a Certified Information Systems Auditor. During his tenure at Zeron, he led the implementation of the Cyber Risk Posture Management Platform with his knowledge and experience in cybersecurity, risk management, information and data security. He holds a Bachelor of Technology degree in Electronics and Communication Engineering from Maulana Abul Kalam Azad University of Technology (MAKAUT). He has a keen interest in Economics involved in Cybersecurity, Shell Scripting, Networking and Risk Management.
\end{IEEEbiography}
% Add other authors here

\end{document}

%% file: sections/introduction.tex
\section{Introduction}
The quantification of cyber risks is critical for companies and governments that want to understand better their exposure to cyber threats \cite{amin2019practical}. This exposure can be technical, where business configurations and assets can increase the likelihood of cyberattacks, but also economical since cyberattacks can cause direct and indirect impacts on business operations that may cause financial losses \cite{economicinnature, rcvar}. Therefore, practical approaches for Cyber Risk Quantification (CRQ) must be considered during different security decision-making processes within companies since it enables businesses to strengthen their overall cybersecurity posture. Examples of scenarios that can benefit from practical CRQ approaches include threat prioritization based on technical and economic risks, measurement of potential losses due to cyberattacks, and strategies to efficiently allocate business resources (\ie money and time) \cite{gartner2023}.

However, CRQ is challenging since quantifying any potential risk a business might face due to cyber threats involves many steps and requires the correlation of different information (from business to technical level) \cite{woods2021sok}. In a recent report \cite{gartner2024}, for example, it was predicted that half of the security leaders would not have success in using CRQ for decision making, against only 36\%, which will effectively be able to reduce risks and operational costs by applying CRQ models. These statistics highlight real-world challenges for companies to apply quantitative methods for efficient risk assessment. Even though there are efforts to process vast amounts of data to support decision-making, it is vital to provide CRQ approaches that add value to a business and provide information that decision-makers need and understand.

For an effective CRQ, different factors must be considered (\eg the likelihood of successful cyberattacks happening and their associated threats), potential impacts of cyberattacks mapped, and cost analysis due to cyberattacks and remediation have to be performed \cite{CyberTEA-Paper}. However, such quantification is challenging since it requires knowledge from a Technical, Economic, and Legal (TEL) perspective regarding a company and the entire cybersecurity landscape. Therefore, approaches for CRQ must handle such challenges and find effective ways to circumvent current limitations companies face, including \1 information asymmetry among companies, \2 miscommunications between board levels, and \3 lack of quantitative mapping between threats and their actual TEL impacts.

There are efforts in academia and industry to bridge this gap. In recent years, different works aimed to predict the risks of a cyber incident happening \cite{von2024quanttm, SecRiskAI-Paper} and estimate its potential losses \cite{rcvar, erola}. Also, cyber insurance companies have applied CRQ to assess the insurability of cyber risks \cite{malavasi2022cyber, insurancerequirements} and for premium calculation \cite{jiang2024cyber}. In the industry, CRQ approaches and tools have been used to measure the effectiveness of security controls and quantify the overall security posture. Examples of well-known CRQ approaches include quantile-based models such as the cyber value-at-risk initially proposed by the World Economic Forum \cite{WEF} and improved by the FAIR institute \cite{FAIR}, thus becoming a \textit{de facto} model for information risk management. However, there is still a need for holistic and practical approaches that consider, from a TEL perspective, the business characteristics, its cyber risks, and potential impacts to provide insightful analysis and recommendations based on real-world data and statistical simulations. 

To bridge this gap, in this article, we propose the Quantified Business Exposure to Risk (\approach) approach, a novel approach for CRQ that considers both technical and economic perspectives of cybersecurity to provide quantitative metrics to inform decision-makers on the actual financial risks that their organizations are exposed to. This includes scores on potential financial loss in case of cyberattacks and an in-depth risk analysis based on the current cybersecurity controls implemented within a company. Such risks are described per business units and segments that compose the organization, including their strategic relevance, from the TEL perspective, for the business operation. Also, \approach{} provides a complete cost analysis for companies based on industry reports, standards, international and national regulations, and well-known guidelines for cybersecurity. Thus, using our proposed CRQ approach, decision-makers can conduct relevant steps of financial risk assessment and strategic planning toward better cybersecurity investments and decisions.

The main contributions of this article can be summarized as follows.\begin{itemize}
    \item The mapping of likelihood and risks of cyberattacks based on data collected using open-source intelligence, industry reports, and feedback from security experts;
    \item The definition of potential TEL impacts based on companies' characteristics and business profiles;
    \item A mathematical model to quantify the financial impacts and potential losses in a business in case of cyberattacks; 
    \item A set of recommendations to reduce the risks while ensuring compliance and cost-benefit investments;
    \item An analysis and discussion on trends and challenges for the wide deployment of CRQ as an enabler for real-world companies in different sectors.
\end{itemize}

The rest of this article is organized as follows. Section II provides a review of the CRQ literature. Section III details the \approach{} and describes its steps. Section IV discusses the challenges and opportunities for CRQ identified in this work, followed by Section V, the conclusions and future work.

%% file: sections/relatedwork.tex
\section{Related Work} % aka Related Work
Cyber Risk Quantification (CRQ) is a specialized risk assessment step, focusing on providing data-driven and explainable indicators for the decision-making process. It provides measurable metrics and results that help decision-makers understand dimensions not available in traditional risk assessment methods. This includes understanding exposure to risks, threats, and financial losses. Therefore, CRQ differs from traditional risk assessment regarding goals, inputs, and communication, such as focusing on quantitative results, incorporating data and analytics in the entire process, and enabling better communication with stakeholders to make informed decisions.

Academia has focused on analyzing cyber risks from different perspectives \cite{CyberTEA-Paper, CyberTEA}, including the technical, economic, and legal impacts. Also, novel approaches have been developed to support measuring and mitigating impacts based on such perspectives. Examples include approaches for business-centric threat quantification using Business Impact Analysis (BIA) \cite{von2024quanttm}, prioritization based on economic aspects\cite{gordon-segmentation, franco2023secadvisor, rosi}, cyber insurance analysis based on CVaR \cite{orlando2021cyber}, and analysis of economic metrics for cost-efficient security decisions \cite{collier2023metrics}. Besides that, there are efforts to verify the actual benefits of technical and statistical indicators to make systems safer \cite{woods2021sok}. Overall, state-of-the-art shows potential to map and understand TEL impacts effectively, but it can be measured when security indicators (\eg threat and exposure) are also considered.

Different industry sectors rely on well-established cyber risk models for risk assessment and management. Some widely adopted models can be considered as \textit{de facto} standards, such as FAIR, ISO 27005, and NIST SP 800-30. However, the different models applied have concerns, especially regarding their usage complexity and the number of parameters required for their usage. Table \ref{tab:Models} compares selected models against the \approach.

\begin{table}[h!]
    \centering
\caption{Comparison of \approach{} against Industry Models}
\label{tab:Models}
   \resizebox{\columnwidth}{!}{%h
   \begin{tabular}{c|c|c|c|c} \hline 
         Model &  Type &  Ease of Use &  \makecell{Number of\\ Parameters} &  Popularity\\ \hline 
         FAIR &  Quantitative&  Complex&  High &  High\\ \hline 
         CyberInsight &  Quantitative &  Moderate&  Moderate&  Moderate\\ \hline 
         NIST SP 800-30 &  Qualitative &  Easy&  Low &  High\\ \hline 
         ISO 27005 &  Qualitative &  Complex &  Moderate &  High\\ \hline 
         OCTAVE &  Qualitative &  Moderate&  Low&  Moderate\\ \hline
         QBER&  Quantitative &  Easy&  Low &  -\\ \hline

    \end{tabular}
    
    }
\end{table}

The Factor Analysis of Information Risk (FAIR) \cite{FAIR} allows for quantifying cyber risk in financial terms. It uses several factors (\eg loss events, exploitability level, and controls implemented) to estimate potential financial losses in case of cyberattacks. Although FAIR is a data-driven model that aims to provide an easy-to-use approach for business leaders, it still requires technical training and considerable effort (money and time) to implement it properly. Also, the business must clearly understand its underlying infrastructure and risk scenarios, which can be challenging for most businesses. 

As another relevant model, CyberInsight \cite{CyberInsight} is based on MITRE ATT\&CK and VERIS framework, which focuses on quantifying an organization's cyber risk posture by considering potential threat actors, vulnerabilities, impacts, and controls implemented. CyberInsight can be helpful in resource allocation decisions and comparison with specific industry sector benchmarks. However, there are still challenges since it requires specific datasets and has limited customization options.

Differently, NIST SP 800-30 \cite{NISTSP} stands as a well-established framework for qualitative assessment of cybersecurity risks. It allows organizations to identify vulnerabilities and impacts on assets and implement efficient security controls. The framework is a good point of entrance for cybersecurity since it is straightforward. However, it needs to consider the financial cost perspective explicitly, thus not being the best choice for planning cybersecurity, which requires, for example, understanding the cost-benefit analysis of investments. 

Similarly to well-known approaches, ISO 27005 is an internationally recognized framework for identifying, analyzing, and evaluating security risks. It is built on the principles of the ISO 27000 series to provide a systematic approach to risk management. It requires a significant investment of time and money to fully understand and implement it effectively. This poses challenges for organizations without technical expertise and a limited budget. Also, there needs to be clear guidance on handling complex and sensitive data, thus requiring organizations to use additional models and standards together with ISO 27005. Although it allows customization for the specific needs of organizations, such customization is challenging and requires high expertise in risk management methodologies.

As another example, the Operationally Critical Threat, Asset, and Vulnerability Evaluation (OCTAVE) \cite{alberts2003introduction} provides a structured approach for identifying critical assets, mapping threats, and evaluating potential impacts. It relies on qualitative categories (\eg low, medium, high) to assess the risk severity. OCTAVE is systematic, adaptable, and does not require too many resources to use, thus making it suitable for organizations that prioritize high-level risks. However, it does not consider explicitly the financial dimensions of impacts and relies on very subjective risk categorization.

Therefore, current quantitative CRQ models, although valuable, have limitations. Established models like FAIR require adequate data, complex calculations, and intensive training. Also, the current CRQ models need to take a lot of other inputs as parameters, including all types of risks, whether internal and external risks of an organization, compliance, regulatory risks, or business risks. \approach{} approach automatically takes - based on a few inputs from users - technical, business, legal, and economic parameters in a straightforward way, which helps to assess cyber risk in more detail and intuitively.

%% file: sections/approach.tex
\section{The \approach{} Approach}
The proposed \approach{} addresses the cyber risk quantification problem by providing a novel approach that explores Open-Source Intelligence (OSINT), cybersecurity economics models, and automated risk analysis techniques to provide insightful information for decision-makers to understand their businesses' risks and associated costs. The approach receives companies' information as input, applies data mapping and economic models, infers data based on statistical analysis and reports, and provides a set of measurable metrics and recommendations for strategic decisions based on quantifying technical and economic risks. Different standards and frameworks are integrated within the proposed approach, such as MITRE ATT\&CK for threats identification \cite{MITRE} and the Secure Controls Framework for (SCF) \cite{SCF} for initial mapping of efficient controls against specific threats and risks.

\begin{figure*}[ht!]  
\centering
\includegraphics[width=0.90\textwidth]{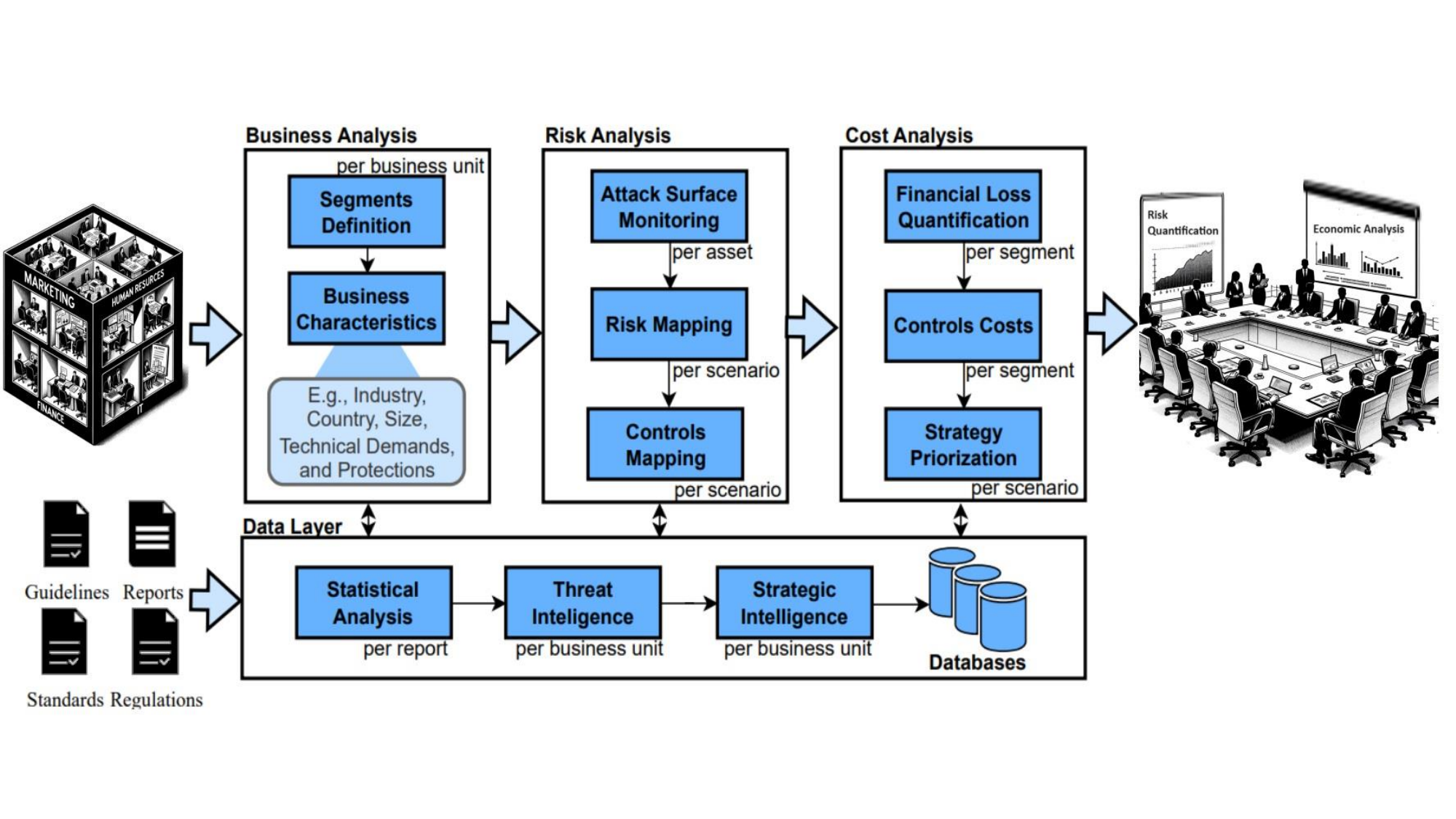}
\caption{The \approach{} Approach}
\label{figure:approach}
\end{figure*}

The overview of the \approach{} is shown in Figure \ref{figure:approach}, including its different modules, components, and their relationships. The \textit{Business Analysis} module receives inputs from the users in order to build a business profile in terms of different business characteristics (\eg country, regulations, sector, and technical demands) and also understand how the businesses' systems and information are structured, such as which segments of information and assets are relevant for the business operation from a technical and economic perspective. Next, the \textit{Risk Analysis} module identifies the risks and controls implemented, including monitoring attack surface based on assets, segments, business units, and associated technical and economic risks. 

Finally, the \textit{Cost Analysis} provides strategic analysis based on economic models. In this module, the potential financial loss is quantified by applying novel and state-of-the-art approaches, the costs to implement controls are measured, and recommendations for cost-efficient cybersecurity strategies are provided. All the modules interact with the \textit{Data Layer} to obtain data and information for the different analyses. This layer includes \1 statistical data extracted from industry reports, standardization documents, and regulations, \2 insightful information regarding potential risks for specific scenarios, and \3 database with controls and guidelines for cost-efficient strategic decisions to reduce risks. 

The output of the \approach{} is a set of quantitative metrics that helps decision-makers to understand technical and economic risks as well as plan and prioritize their cybersecurity strategy. This includes the level of risks for each business unit, segment, and asset. Potential economic losses are also mapped, and risk prioritization can be performed to reduce financial impacts while ensuring compliance with different guidelines and regulations. Each of the modules and their relevant steps are described in detail in the rest of this section. The approach and its metrics have been applied as part of commercial products for the Banking, Financial Services, and Insurance (BFSI) sectors\footnote{https://zeron.one/}.

\subsection{Business Analysis}
% How we do build the business profile and why this information is required.
The business analysis is conducted automatically and manually by collecting information regarding the business. Initially, the business details, which most companies know, are required as input, such as the size of the business revenue, number of employees, country of operation, sector, and different business units that compose the company. Next, the specific segments worthy of each business unit are defined. For example, a database of customers and a payment system can be more critical for a specific business unit than others, while web servers are the most important for other business units that offer specific services.

Note that we consider the following definitions. Business unit is a division within a company responsible for specific services with its management, resources, and security. Segments are specific information or services within business units that have different values for the business unit operation. Together, business units and segments have risks that impact with different magnitude the entire company of which they are part. Therefore, it is necessary to consider such a fine-grained analysis when collecting and understanding the business profile.

\begin{table}[h!]
  \centering
      \caption{Overview of Information Collected to Build the Business Profile}
      \label{table:business}
\resizebox{\columnwidth}{!}{%h
 \begin{tabular}{c|c|c} 
 \hline 
 \textbf{\makecell{Metric}} & \textbf{\makecell{Description}} & \textbf{\makecell{Example}} \\
 \hline
 \makecell{Business \\ Size} & \makecell{Business revenue, \\employees,\\ and business units} & \makecell{Medium-sized business  with \\ \$ 10 M as yearly revenue \\ and 4 business units} \\
 \hline
 Sector & \makecell{Industry sector where \\ the business units operate} & \makecell{Banking, financial \\ and insurance} \\
 \hline
 Country & \makecell{Country sector where \\ the business units operate \\ or offer services} & \makecell{India,\\ United States} \\
 \hline
 \makecell{Segments \\ and Services} & \makecell{Segments of information \\ and services that have \\ economic value for the business} & \makecell{Customer Data, \\Payment System, \\ Sales platform} \\
 \hline
 \makecell{Economic \\ Details} & \makecell{Profits and revenue per \\ business units and segments} & \makecell{60\% of revenue from \\ the sales platform} \\
 \hline
 \makecell{Implemented \\ Controls} & \makecell{Controls already Implemented \\ and its current maturity} & \makecell{Access Control, WAF, \\ Endpoint Protection} \\
 \hline
 \end{tabular}
 }
\end{table}

\begin{figure*}[ht!]  
\centering
\includegraphics[width=0.80\textwidth]{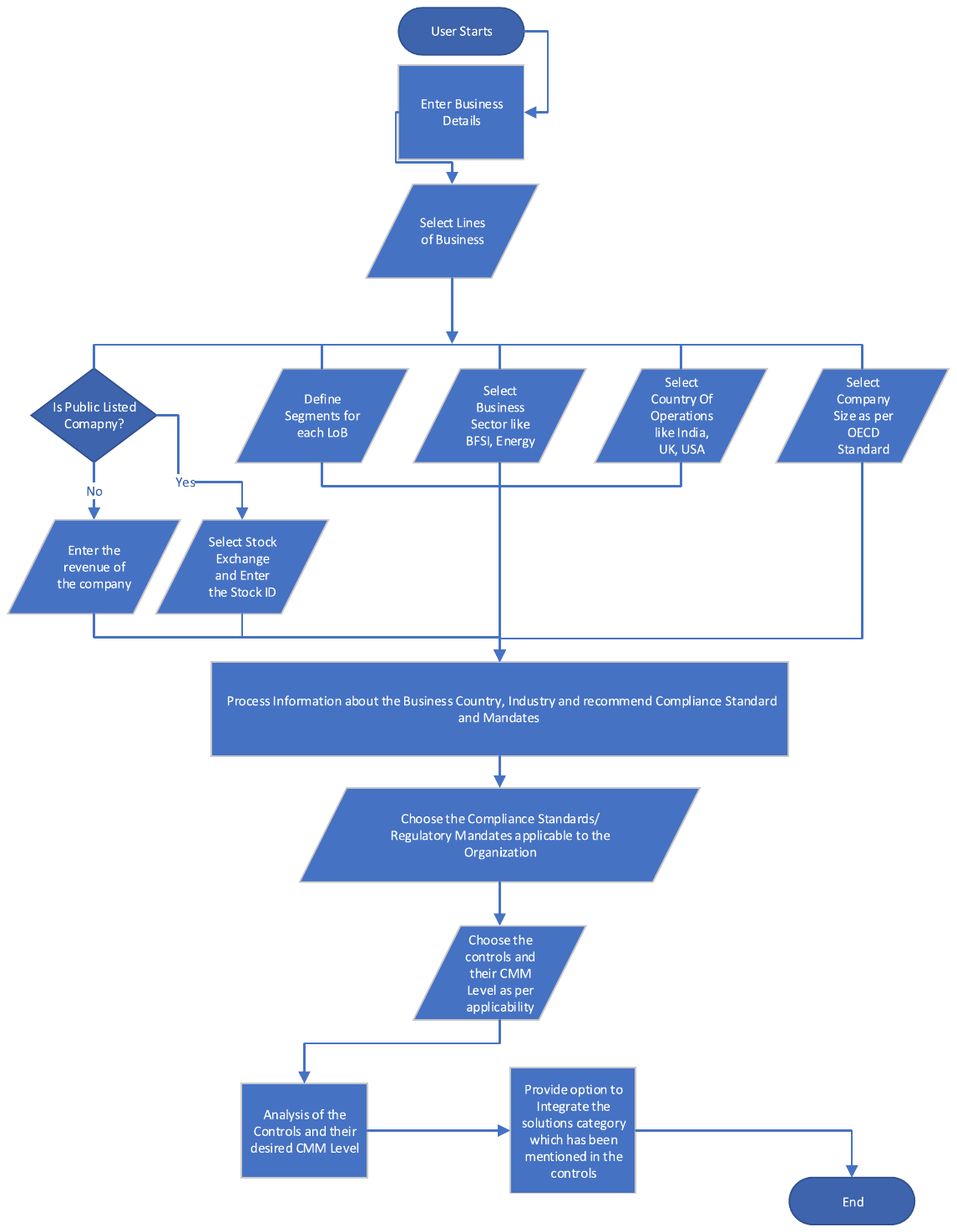}
\caption{User Flow of \approach{}'s Information Collection for Business Analysis Implemented in a Commercial Solution¹}
\label{figure:flow}
\end{figure*}

Table \ref{table:business} highlights the general information collected during the business analysis. This also enables the companies to select specific compliance standards applicable to business units and define which controls are already in place to protect each business unit and specific segments. Understanding the importance and how worthy (percentage of revenue) a business unit and segment for the company is challenging but a key for accurately quantifying TEL impacts. A guided approach is provided to the companies to collect all data needed to infer the importance and relevance of such critical economic details. Figure \ref{figure:flow} shows the user's flow of information and examples of paths and information provided in whether publicly listed companies are being considered. All these steps are performed using intuitive and guided interfaces suitable for decision-makers.

As risk factors and TEL impacts may differ according to the businesses' characteristics, providing an in-depth analysis and accurate information is essential. If information is missing, the \approach{} can use information obtained from OSINT, industry reports, or even penetration tests and security analysis results. Also, market averages, stock exchange analysis, and surveys with similar companies can be used as input to understand the actual market trends. Thus, we provide a database containing the mapping of business profiles with threats, risks, and TEL impacts. This is used as input for the risk analysis module.
% Table with general information we collect to understand the business.

\subsection{Risk Analysis}
% Write the intro and create the links
After collecting and processing the relevant information provided by the company, the \approach{} starts the risk analysis process. This starts by automatically analyzing all business units and assets to understand the attack surface (\eg monitoring assets, segments, and business units as seen by potential attackers). This is conducted using proprietary solutions that provide a technical analysis highlighting the most vulnerable points, such as assets with a higher likelihood of an attack due to many Common Vulnerabilities and Exposures (CVEs). 

Although such a step is relevant for a precise CRQ, it is not easily interpreted by the company's board. Therefore, after having all the technical details, it is necessary to understand and quantify the potential risks from a TEL perspective. A rich database maps threats, controls, and associated costs. Such a database is defined following \textit{de facto} standards (\eg NIST \cite{NIST}, SCF \cite{SCF}, and ISO 27000 series \cite{ISO27001}), security experts' opinions, and data collected using OSINT techniques from real-world and also industry reports \cite{rcvar}. By using such information combined with the company details, the \approach{} can conduct a detailed analysis to quantify not only risks and potential impacts but also the potential financial losses in different scenarios (\eg based on information mapping, Monte Carlo simulations, and Cyber-Value-at-Risk models).

Table \ref{table:terminology} summarizes the terminology used for the calculations. All symbols are appropriately represented, including their definitions and descriptions for each. The different metrics computed as part of the risk analysis module of the \approach{} are introduced below.

\begin{table}[h!]
  \centering
      \caption{Terminology of Symbols and Metrics Used on Equations}
      \label{table:terminology}
\resizebox{\columnwidth}{!}{%h
 \begin{tabular}{c|c|c} 
 \hline 
 \textbf{\makecell{Symbol}} & \textbf{\makecell{Metric}} & \textbf{\makecell{Description}} \\
 \hline
 $w$ & Weight & \makecell{Value from 0 to 10 \\ that describes \\ Low, Medium, or \\ High weight for\\ a given metric} \\
 \hline
 $Seg$ & Segment & \makecell{Segment of information or\\ service that has \\ value for a business} \\
 \hline
 $Impacts$ & \makecell{Operational \\ or Financial \\ impacts} & \makecell{Impacts measured from a\\  technical and economic \\perspective} \\
 \hline
 $T$ & Threat & \makecell{Threat under analysis, \\ such as Phishing,\\ Malware, and DDoS} \\
 \hline
 $D$ & Domain & \makecell{Represent the different ways \\and dimensions that a \\cybersecurity strategy can consider} \\
 \hline
 $Risk$ & \makecell{Technical \\or Economic \\ risk} & \makecell{The exposition to an cyberattack \\ considering likelihood and\\ potential success rate}\\
 \hline
 $CIA$ & \makecell{Confidentiality,\\ Integrity,\\ Availability} & \makecell{Analysis of security posture\\  based on the CIA triad} \\
 \hline
 $\alpha$ & Coefficient & \makecell{Value used to \\ normalize the results\\ of equations} \\

 \hline
 \end{tabular}
}
\end{table}

% Impacts - Operational and Financial Impact

% Seg - Segment
% w - Weight
% T - Threat
% D - Domain
% \alpha - Coefficient

% Impacts
The \approach{} initially considers two main impacts: Operational and Financial. The operational reflects the capacity of a cyberattack to cause service disruption or infrastructure damages, while the financial impact maps the magnitude of potential financial loss due to a cyberattack. Therefore, the impacts of a potential threat ($Impact_{w}$) are given as an operational and financial impact factor. For example, for a given business unit that handles payments, in the case of data leakage due to a cyberattack, the operational impact can be \textit{Low or Medium} if the data is not encrypted but only leaked. In contrast, the financial impact will be \textit{High} due to regulatory and compliance costs. 

The operational impact might also be \textit{High} if Ransomware disrupts the entire payment service. The final impact will be generated based on potential operational and financial impacts. Equation \ref{equation:Impacts} shows the impact calculation per threat. The values for each threat must be mapped according to the known potential impacts based on the analysis of the business profile, characteristics, and market trends.

\begin{equation}
    \label{equation:Impacts}
    \begin{aligned}
       Impacts_{w} = Impact_{Operational} * Impact_{Financial} \\
        \end{aligned}
\end{equation}

% Risk Score
Next, the \textit{Economic Risk Score (RS)} is determined based on cybersecurity economic models available in the literature and technical metrics collected from the business. For that, the RCVaR model \cite{rcvar} is used since it provides a view of the different characteristics of a business and factors that can increase or decrease the risk of economic impacts in case of cyberattacks. To provide such analysis, RCVaR uses statistical analysis of different industry reports and provides. RCVaR is then integrated into the \approach{} to quantify the economic impacts and also determine the magnitude of potential impacts (\textit{low, medium, or high}). 

Equation \ref{equation:EconomicScore} defines the economic risk score. This is achieved by computing the average of RCVaR factors times the current cybersecurity posture implemented regarding Confidentiality, Integrity, and Availability (CIA). Note that the CIA is obtained by analyzing the current controls implemented by the company and how they help to ensure the different elements of the cybersecurity triad. The $RS_{Economic}$ metric provides then an analysis of the magnitude of exposition to economic impacts due to cyberattacks. The potential economic impact is computed later using additional metrics (\cf Section \ref{section:costanalysis})

\begin{equation}
    \label{equation:EconomicScore}
    \begin{aligned}
       RS_{Economic} = avg(RCVaR_{Factors}) * avg(CIA) \\
        \end{aligned}
\end{equation}

Based on the risks identified and the business profile, decision-makers can prioritize where to invest their time and money to reduce them. For that, the \approach{} highlights, using the concept of domain prioritization, which kind of protection can be prioritized against specific cyberattacks.

Domain prioritization ($D_priority$) involves determining which strategy might be applied first to reduce the potential impacts of specific threats. For example, the domains \textit{People and Awareness and Endpoint Protection} must have prioritization in case of planning against \textit{Phishing} attacks. At the same time, Endpoint and Database protection can be considered for Ransomware cases.

Therefore, it is possible to determine in which way and dimension that company must focus resources when defining a cybersecurity strategy. Equation \label{equation:Domain} defines how the prioritization is calculated, where $T_{w}$ represents how relevant a threat is for a domain and $Impact_{w}$ determines the magnitude of TEL impacts in case of $T$ happen. \approach{} considers fifteen cybersecurity domains extracted from technical standards.

% Domain prioritization
\begin{equation}
    \label{equation:Domain}
    \begin{aligned}
       D_{priority} = T_{w} * (\alpha + Impact_{Weight}) \\
        \end{aligned}
\end{equation}

% Attack surface and logs

% MITRE? Risk score? Z-score?

% Create the link between sections
\subsection{Cost Analysis}
\label{section:costanalysis}

By knowing the risks the company might face and the magnitude of potential impacts, the \approach{} provides a quantification of the financial losses that the company can expect per asset and segment of information. This is calculated considering the monetary value of a given segment for the business and impacts previously collected from industry reports and mapped in our database (\eg the potential operational and economic impacts of a DDoS in an e-commerce). These impacts are determined as \textit{high, medium, and low} and later converted to a range of numbers for the segment impact quantification ($Seg_{Impact}$).
% Database modeling

Equation \ref{equation:SP} shows how the financial impact is calculated per segment. It is considered the revenue of the business that is dependent on a specific segment or asset, and then this monetary value is adjusted based on the impact of a specific threat. Thus, the $Seg_{Impact}$ should be computed per segment and threat. This can be used as an independent value to understand the economic risks of specific threats (\eg Phishing or Ransomware) to a given segment. However, it also can be used as a weighted quantification to determine the overall economic impact (considering all threats).

% Segment Impactact
% SP = Seg_revenue * (OI * EI)
\begin{equation}
    \label{equation:SP}
    \begin{aligned}
       Seg_{Impact} = Seg_{Revenue} * Impact_{w} \\
        \end{aligned}
\end{equation}

After the definition of the potential impact (worst scenario), it is computed the actual segment risk ($Seg_{Risk}$) by considering the impact's magnitude and also the risk of an attack being successful. Therefore, $Seg_{Risk}$ is the actual risk based on the current controls implemented and the likelihood of an attack happening and being successful. Equation \ref{equation:SR} computes its risk and provides, as output, the monetary value quantified as possible loss in case of a cyberattack in such a segment. If new controls are implemented, the $Seg_{Risk}$ can be computed again to verify the updated economic risk score.
% SE = SP * (Risk * Vulnerability)

\begin{equation}
    \label{equation:SR}
    \begin{aligned}
       Seg_{Risk} = Seg_{Impact} * (Impact_{w} * Risk_{w}) \\
        \end{aligned}
\end{equation}

The effectiveness of controls against specific threats has also to be considered. This effectiveness is defined based on its capacity to reduce the potential risk of successful attacks and TEL impacts. The control costs can also be compared against others since controls require high costs due to associated Capital Expenditures (CAPEX) and Operational Expenditures (OPEX). On the other hand, solutions with reduced CAPEX and OPEX might slightly reduce the risks (even in a lower magnitude) but with a better cost-benefit. 

Note that the maturity of the controls implemented is also crucial since the the effectiveness of a control is driven by how it is implemented \cite{Effectiveness}, which includes security policies and controls' maturity. Therefore, A company with initial or repeatable maturity might have an adjustment on the actual control efficacy, while an optimized maturity ensures the best case of control efficacy. This is computed as part of our model to find a company's actual protection (and risks) based on the implemented controls. The \approach{} assigns a quantitative score to each control ($Maturity_{Level}$) according to its maturity and uses it to reduce or increase the original efficacy of a control based on its maturity level:
\begin{itemize}
\item Not Implemented: 1.25 
\item Initial: 0.65 
\item Repeatable: 0.55 
\item Defined: 0.45 
\item Managed: 0.31 
\item Optimized: 0.25 
\end{itemize}

These values are used together with a control importance score to determine the actual efficacy of the control. For example, suppose the maturity of control is initial, and the control importance score is \textit{high}. In that case, it means that the efficacy of the control would be reduced from \textit{high} to \textit{medium}. Therefore, this means that controls with high importance and efficacy might have the efficacy reduced in case the maturity of the controls is not adequate. The mapping of maturity and its importance score is done as follows. The actual control efficacy ($ControlEfficacy_{T}$) for a given threat is calculated using the promised efficacy and the current maturity level implemented, as shown in Equation \ref{equation:controlefficacy}. As input for the $Efficacy_{T}$. \approach{} provides a map of control's importance based on different risk scenarios. 

\begin{equation}
    \label{equation:controlefficacy}
ControlEfficacy_{T} = Efficacy_{T} * (1.25 -Maturity_{Level})
\end{equation}

% Cost-benefit of investments - Modified ROSI
A set of recommendations can be provided based on compliance needs. However, companies tend to have a limited cybersecurity budget and must decide on the most cost-effective protections while encompassing compliance. For that, a modified version of the well-known Return on Security Investment (ROSI) model \cite{rosi} is provided. The modified ROSI named as Z-ROSI is shown in Equation \ref{equation:Z-ROSI}. As can be seen, it differs from the original ROSI by integrating the cost rate and technical exposure into the calculation. This creates a specialized model that can be adapted according to the company's demands, even for companies that do not have a clear view of their risks, actual technical and financial exposure, and control costs.

% Equation modified ROSI

\begin{equation}
    \label{equation:Z-ROSI}
\resizebox{0.5\textwidth}{!}{$Z\text{-}ROSI = \frac{((ALE * Control_{Efficacy}) - (Control_{Cost} * Cost_{Rate})}{(Control_{Cost} * Cost_{Rate})}$ \\
}
\end{equation}

In Equation \ref{equation:exposure}, the exposure is computed as a function of the CIA obtained by previous analysis. This shows the percentage of exposure of the company to CIA faults. Next, in Equation \ref{equation:ALE}, the Annual Loss Expectancy (ALE) is defined based on the calculated exposure and the current potential financial loss of a given segment or business unit.

% ALE and exposure equation
\begin{equation}
    \label{equation:exposure}
    \begin{aligned}
       Exposure = 1 - avg(CIA) \\
        \end{aligned}
\end{equation}

\begin{equation}
    \label{equation:ALE}
    \begin{aligned}
       ALE =  Exposure * Seg_{Impact} \\
        \end{aligned}
\end{equation}

Thus, applying the Z-ROSI makes it possible to understand if a specific control to be implemented is cost-effective. For that, it considers the cost of the control, the current potential loss of the asset that the control will protect, and the risk exposure. This ensures a more realistic and suitable approach for ROSI calculation since it uses information collected from the company using the \approach{} and infers complex information that usually is not entirely known by companies, such as risks and potential costs.

%May add here a table summarizing the KPIs we calculate? KPI -> description -> Equation number. This can be done based on the document we did internally.
% In summary, the \approach{} approach provides [...]. Table X shows a list of Key Performance Indicators (KPIs) that are provided by applying the models and metrics proposed for the business, risk, and cost analysis.

In summary, the \approach{} provides a well-defined approach and model for quantifying risks from an economic and technical perspective. The model comprises a set of metrics that helps decision-makers understand their business' cybersecurity posture and better plan their cybersecurity strategies. The metrics are provided for a complete analysis and quantification of how much the business is exposed to financial impacts due to cyberattacks. They also highlight potential weaknesses in the business' cybersecurity strategy that must be prioritized. 

Thus, using the \approach, it is possible to identify loss probability trends and risk scenarios to specific threats, prioritize investment based on identified risks, and understand the efficacy of current controls implemented within different business units. Also, the cost of investing in cybersecurity is covered by applying economic metrics that help to find cost-effective controls to reduce potential TEL impacts and overall risk exposure.

%% file: sections/trends.tex
\section{Demands, Trends, and Challenges for CRQ}
% Demands starting here
% Compliance, 
% Industry needs
CRQ is gaining attention mainly due to the demands of companies from different sectors that urge them to make cybersecurity decisions effectively. Influential in this context means protecting their main assets and encompassing regulations without overspending their budget. For that, it is critical to have more data and metrics and insightful information that can be used by multidisciplinary stakeholders (\eg from business, cybersecurity, statistics, and economics fields) with different levels of expertise. Therefore, this is a demand from industry and society to take control of their posture to survive in the digital world. 

% Gartner trends
In a recent survey conducted by Gartner \cite{gartner2023} with board managers and the C-suite, it was observed that CRQ has been used chiefly for cyber insurance and compliance reporting, followed by prioritization of security investments. CRQ models increase the confidence of the board and make it easier to convince risk owners to remediate risks. Also, the business exposure can be better understood during risk analysis tasks. We can see an increasing investment in CRQ  as a trend. However, stakeholders still need help understanding the CRQ analysis due to the lack of guidance and the subjective nature of current CRQ methodologies.

% Cyber Insurance trends
Cyber insurance companies are also relying on CRQ models \cite{orlando2021cyber, palsson2020analysis} to provide better premium calculations, coverage, and risk appetite analysis \cite{MALAVASI202290}. According to a recent report from Munich RE \cite{munichre}, one of the biggest world cyber insurance companies, the finance sector is responsible for more than half of all cyber insurance claims, with ransomware, business communication compromise, and data breaches being the major loss drivers. It is also possible to see a trend for AI-based protections being used, with future AI regulations having the potential to impact the compliance needs of companies in terms of cybersecurity.

From the cyber risk perspective, AI-based solutions \cite{SecRiskAI-Paper, sarker2021ai} have arisen as a potential ally to address the asymmetry of information issues and also to identify patterns in attack behaviors and their impacts. For the prioritization of threats, the Exploit Prediction Scoring System (EPSS) \cite{jacobs2021exploit} has gained attention from the industry due to its capacity to estimate the probability of a vulnerability being exploited in the next 30 days. The EPSS relies on real-world exploit and threat information from many OSINT and proprietary sources. % challenge: frequent data updates that may change dramatically the security strategy. Technical-oriented.

The economic impacts of cyberattacks still lack a \textit{de facto} model. The companies tend to rely on general models like the Business Impact Analysis (BIA) \cite{von2024quanttm} and FAIR model \cite{FAIR} to understand potential impacts. However, these models still depend on subjective elements that may vary according to their applications and configurations. Companies also adopt economic models. Examples are the well-known ROSI \cite{rosi} for calculation of cost-benefit of cybersecurity investments and other classical models like CVaR for estimation of financial losses \cite{orlando2021cyber} and variations of Gordon-Loeb \cite{gordon-segmentation, gordon-insurance} for optimal cybersecurity investment. From the academia, there are also prominent applications using the CVaR and Gordon-Loeb as a basis, such as the RCVaR model \cite{rcvar} that provides insights on financial loss based on industry reports and SECAdvisor \cite{franco2023secadvisor} that applies the Gordon-Loeb model as part of the cybersecurity planning. 

%  Challenges starting here.

% drawbacks of current models in terms of complexity
Although certain models have gained traction in the cybersecurity community, there are still drawbacks that have to be highlighted and further addressed. Most models are resource-intensive since they need substantial industry data, time, and technical expertise for comprehensive risk analysis. Also, the current models rely on probability estimation, which may foster a false sense of certainty, thus potentially undermining risk assessments. These limitations can be seen in widely used models, such as the FAIR model and Gordon-Loeb.

% threar priorization
The need for vulnerability prioritization and strategic reasoning is also a limitation. This has been a concern for both academia and industry, which are focusing on providing ways to understand what has to be solved now and what can wait. Therefore, one of the trends mapped is methodologies and models for prioritizing vulnerabilities, resources, and monetary investment. Examples include the EPSS or even the usage of EPSS and CVSS together, as recommended by FIRST \cite{FIRST-EPSS}. However, there are still challenges, especially due to the dynamic nature of data that may change security strategies substantially every day. Also, such methodologies tend to be very technical-oriented, thus making it hard to be used to draw the attention of the board and C-level within companies.

The efficacy of controls is related not only to whether the control is implemented but also to how the control is implemented. Therefore, it needs to understand the business processes and the maturity of the implemented processes and controls. An in-depth analysis is challenging because many business and technical elements must be checked when understanding the maturity. However, there are established solutions - as applied by the \approach{} approach - that can help to understand the maturity to be used as part of more complex models. Examples of such solutions include the IT General Controls Capability Maturity Model (CMM) \cite{CMM} and the SCF \cite{SCF}. Automated and AI-based approaches can support data collection to analyze business characteristics and attack surfaces better. However, AI may still face challenges regarding adversarial attacks and regulations.

Furthermore, the industry is moving towards a more explainable cybersecurity cost analysis, while academia has also focused on providing more accurate and intuitive models. More intuitive and efficient CRQ becomes an urgent need for companies since cybersecurity is becoming too complex to operate and understand, with vast amounts of data and technical metrics being provided daily for a - still challenging - decision-making process. Approaches like the proposed \approach{} are utmost to empower companies with well-defined steps - anchored by real-world data and business demands - for quantification of risks and associated costs.

%% file: sections/conclusions.tex
\section{Conclusions and Final Remarks}
% CRQ is key for companies
CRQ is critical for companies to survive in a digital world. Cyberattacks continue to increase risks and TEL impacts while efficient cybersecurity strategies are in demand. Determining an efficient strategy requires understanding information in a qualitative and quantitative way, thus empowering decision-makers with more data and measurable metrics to guide the decision process. For that, CRQ approaches can benefit technical and non-technical people in determining what to prioritize, where to focus, and how to protect.
% Our work address such issue by doing this and this

The \approach{} approach is proposed to address such scenarios where measurable metrics must be provided clearly in a way to make informed decisions. For that, \approach{} proposes a model to quantify actual financial risks the organizations are exposed to while also providing insights about possible legal (\eg compliance to specific regulations) and technical risks (\eg level of controls implemented and potential threats). The likelihood and risks of cyberattacks are provided based on the mapping of real-world data and security experts' analysis, followed by the definition of potential impacts based on companies' characteristics and business profiles. A model is proposed to compute the actual risks and potential financial impacts in case of cyberattacks, followed by recommendations to reduce the risks cost-effectively. \approach{} relies on extensions of well-established models and provides novel equations to quantify such risks and impacts.

% Discussions show trends to this and highlight challenges
The analysis of the threat landscape and current efforts for CRQ shows that it is gaining prominence, particularly for cyber insurance, compliance reporting, and security investment prioritization. It enhances board confidence and facilitates risk mitigation efforts. Despite increasing investment, stakeholders need help understanding CRQ analysis due to its subjective nature and lack of guidance. It shows the need for explainable and easy to use approaches like \approach{} to guide the CRQ process and provide a complete approach containing a flexible model (\eg addressing different scenarios, regulations, and controls) and extensible datasets for accuracy calibration.

% Future directions include
Future work includes validating the proposed approach in selected real-world companies and security experts. Also, a quantitative analysis is mapped to compare our mathematical models' behavior against others in literature (\eg Gordon-Loeb, Monte Carlo simulations, FAIR, and ROSI). Further, an in-depth analysis of the accuracy of the estimations and recommendations provided by the \approach{} approach will be provided. Along with this, more data points from different Security Solution would be supported which would help in enhancing the analysis of risks. Such evaluations can validate our approach as a key for cybersecurity posture management and strategic planning.